\documentclass[12pt]{article}
\usepackage{makeidx}
\usepackage{epsfig}
\usepackage{mathrsfs}
\usepackage{latexsym}
\usepackage{amsfonts}
\usepackage{amssymb}
\usepackage{amsmath}
\usepackage{amsthm}
\def\be{\begin{equation}}
\def\ee{\end{equation}}
\def\bea{\begin{eqnarray}}
\def\eea{\end{eqnarray}}
\begin{document}
\title{
{\bf  Nonperturbative SL(2,Z) (p,q)-strings manifestly realized on the quantum M2}
}
\author{
\hspace*{-20pt} \small \bf M.P. Garcia del
Moral$^1$, I. Mart\'\i n
$^2$, A.
Restuccia$^{2,3}$\\
{\small $^1$ Dipartimento di Fisica Teorica, Universit\`a di Torino}\\[-6pt]
{\small and INFN - Sezione di Torino; Via P. Giuria 1; I-10125 Torino, Italy}\\
{\small $^2$ Departamento de F\'\i sica, Universidad Sim\'on Bol\'\i var}\\[-6pt]
{\small Apartado 89000, Caracas 1080-A, Venezuela} \\
{\small $^3$ Max-Planck-Institut f\"ur Gravitationphysik, Albert-Einstein-Institut}\\[-6pt]
{\small and M\"ulenberg 1, D-14476 Potsdam, Germany}\\
}
\maketitle

 \ \vskip -5in {\em \
%\parbox[t]{1.1\linewidth}
{\rightline{DFTT-29/2008}\\ \rightline{\ \ AEI-2008-002} } } \vskip
4in
\begin{abstract}
The $SL(2,Z)$ duality symmetry of $IIB$ superstring is naturally
realized on the $D=11$ supermembrane restricted to have
 central charges arising from a nontrivial wrapping. This supermembrane is
minimally immersed on the target space (MIM2). The hamiltonian of
the MIM2 has a discrete quantum spectrum. It is manifestly invariant
under the $SL(2,Z)$ symmetry associated to the conformal symmetry on
the base manifold and under a $SL(2,Z)$ symmetry on the moduli of
the target space. The mass contribution of the string states on the
MIM2 is obtained by freezing the remaining degrees of freedom. It
exactly agrees with the perturbative spectrum of the $(p,q)$ $IIB$
and $IIA$ superstring compactified on a circle. We also construct a
MIM2 in terms of a dual target space, then  a $(p,q)$ set of
non-perturbative states associated
to the $IIA$ superstring is obtained.
\end{abstract}
\clearpage
\tableofcontents

\section{Introduction}
In 1981 the $SL(2,R)$ symmetry of the $IIB$ supergravity was found
\cite{julia}. A discrete subgroup of this symmetry $SL(2,Z)$ is
conjectured to be the exact symmetry of the full quantum type IIB
theory, \cite{hulltownsend}. In fact this relation between classical
and quantum symmetries seems to be a generic one
\cite{hulltownsend}. A circumstance that gave strength to this
argument is the fact that the string coupling constant transform in
a no-trivial way under this symmetry.

In \cite{schwarz}  a major step towards understanding this symmetry
was obtained at the level of supergravity analysis of type II
strings. A $SL(2,Z)$ multiplet of string solutions of type $IIB$
theory were found. It was also argue that this symmetry should have
an origin in the 11D supermembrane.  These solutions are called
(p,q)-strings, where $p,q$ corresponds to coprime integers
associated to the electrical charge of RR and NSNS 3-form field
strengths, $F_{3}, H_{3}$ respectively. These solutions are stable
\cite{schw2}, they do not decay into other string states since it
would violate charge conservation. Moreover, the tension of the
bound state is less than the tension of two individuals strings.
This (p,q) strings may be seen as bound-states between fundamental
string and D-string \cite{witten}. The elementary type $IIB$ superstring
 is a source of the NSNS B-field but not of the RR fields \cite{witten}.
 It corresponds to a $(1,0)$ string and through a $SL(2,Z)$ can be mapped
 to a generic $(p,q)$ string. On the other hand a D-string has charges
 $(0,1)$ since carry RR charge but not NSNS one. The $(NS5,D5)$ also
form a $SL(2,Z)$ multiplet with the NSNS 5-form \cite{boonstra}.
 An exhaustive study of these  $SL(2,Z)$ bound-states
 in 9D supergravity was done in \cite{tomas}.
 These results were
 extended for
$D<10$ by \cite{roy}\cite{Lu}, and the explicit tensions and charges
of more general bound states related through T-duality, $(F1,Dp)$
and $(NS5, Dp)$ with $p$ even and odd were explicitly computed
\cite{jabbari}. Since these string states are BPS saturated, this
gave strong evidence to think that the compactified theory also has
an exact $SL(2,Z)$ invariance. The qualitative results are seen to
be preserved. In less than $D<10$ the symmetry group is enhanced and
conjectured to be the U-group which contains the T-symmetry group
and the S-duality \cite{hulltownsend}. Due to the fact that this
symmetry is nonperturbative not many results have been obtained so
far. In the case of the IIA theory these bound states do not appear
at the level of the perturbative spectrum. Since IIA and IIB
theories when compactified on a circle share the same massless
spectrum (although they are different theories as pointed out in
\cite{abou}), it was conjectured by \cite{witten} and
\cite{hulltownsend} that in 10D an array of D0 branes should appear
non perturbatively carrying the information of those  $D=9$ $(p,q)$
states. We will report about
 these states along this paper.\newline

The origin of a 12 dimensional IIB theory in terms of an eleven
dimensional theory M-theory at first sight can be surprising. In
 \cite{azcarraga},\cite{townsend} this aspect was analyzed.
They found that the 12th dimension, instead of a time coordinate,
could admit an interpretation as the M2
tension, which can be regarded as the flux of a three-form
worldvolume field strength. In \cite{abou} the precise meaning of T
duality between $IIA$ and $IIB$ superstring theories compactified on
circles of radius $R$ and $1/R$ was emphasized by considering $N=2$
space-time SUSY in nine dimensions. Moreover duality properties
between M-theory compactified on $T^{2}$ and $IIB$ on $S^{1}$ were
derived from $N=2$ $D=9$ supergravity. These duality properties are
in complete agreement with the more general analysis we will present
in terms of the $D=11$ supermembrane.
Also in \cite{uehara, uehara2} the identification of the $\tau$ parameter
 and the moduli of the $(p,q)$ string tension was obtained. \newline

In this paper we consider the $D=11$ supermembrane restricted by a
topological condition which may be interpreted as if the
supermembrane has a central charge generated by an irreducible
winding (M2MI). The theory is completely consistent since the
topological constraint commutes with all constraints of the $D=11$
supermembrane, it determines a very special harmonic map from the
base manifold to the target. It is an holomorphic map and hence a
minimal immersion of the base manifold onto the compactified sector
of the target. The main property of the MIM2 is that its hamiltonian
has a discrete spectrum. It does allow a direct analysis of its
symmetries at the quantum level. These symmetries which we will
determine explicitly are related as expected to the $SL(2,Z)$ $IIB$
duality symmetry. In particular they ensure that some assumptions in
\cite{schwarz} concerning the supermembrane winding modes are valid
as well as explain why the supermembrane can be wrapped on a torus
any number of times but only one on a circle. Moreover by
considering appropriate matching-level conditions on the MIM2 and by
freezing pure membrane states we can explicitly obtain all $(p,q)$
string configurations within the MIM2. We will obtain the complete
$mass^{2}$ contributions of those configurations and they match
exactly with the $(p,q)$ analysis in \cite{schwarz}. We will also
consider some decompactifications limit of interest,and introduce
the T-dual MIM2 theory. From it we will obtain a set of $(p,q)$
multiplet of non-perturbative states dual to the $IIB$ $(p,q)$
states corresponding to $IIA$ theory.

%%%%

The paper outline is the following: In section 2 we will briefly
summarized the properties of the MIM2, a sector of the theory
corresponding to nontrivial irreducible wrapping of the M2. In
section 3, we show that the MIM2 on a $T^{2}$ posseses two different
$SL(2,Z)$ symmetries,
 one associated to the conformal symmetries of the Riemann basis
 and other associated to a
symmetry of the target space. The simultaneous action of these two
symmetries generate a transformation rule for the axion-dilaton of
the IIB, the radius of the target space and the gauge fields. In
section 4 we reproduce the perturbative mass spectrum
 of the IIB theory covariant under the
$SL(2,Z)$ and $IIA$ when compactified on a circle. In section 5 we
detail the string states of the MIM2 obtaining also the natural candidates to
nonperturbative (p,q)-string states of the IIA conjectured long time ago.
We show the decompactification limits in which ordinary type IIA and
type IIB strings are obtained. Finally we clarify an open puzzle
with respect to the difference between compactifying the M2 on a
circle (which can be consistently made only for winding  $n_{1}=1$)
and compactifying on a torus with an arbitrary wrapping $n$.  In
section 6 we summarize and discuss the main properties of these
results and conclude.

\section{The $D=11$ Supermembrane restricted \- by the topological constraint
 (MIM2).}

The $D=11$ Supermembrane action \cite{bst} is \bea\label{1}
S=-T\int_{M_{3}}d\xi^{3}[\frac{1}{2}\sqrt{-g}g^{ij}\Pi_{i}^{M}\Pi_{jM}
-\frac{1}{2}\sqrt{-g}+\epsilon^{ijk}\Pi_{i}^{A}\Pi_{j}^{B}\Pi_{k}^{C}B_{CBA}]
\eea where $A=(M,\alpha)$ and \bea
&\Pi_{i}^{M}=\partial_{i}X^{M}-i\overline{\Psi}\Gamma^{M}\partial_{i}\Psi
\\ \nonumber
&\Pi_{i}^{\alpha}=\partial_{i}\Psi^{\alpha}\eea

$X^{M}, M=0,\dots,10$, and $\Psi$, a Majorana spinor on the target space,
 are scalars under diffeomorphisms on $M_{3}$. \newline
We will consider $M_{3}=R\times \Sigma$, $\Sigma$ a compact Riemann
surface of genus $g$. The local coordinates are denoted $\tau$ , a
time coordinate on $R$, and $\sigma^{a}, a=1,2$, on $\Sigma$.  The
action is invariant under diffeomorphisms on $M_{3}$, space-time
supersymmetry and $k$-symmetry. The phase space is restricted by
first class constraints generating the above symmetries and by
fermionic second class constraints. We will consider, in this work,
the target space to be a 9 dim Minkowski space $M_{9}$ times a 2-dim
flat torus $T^{2}$. An extension to other compactified sectors on
the target space were considered in
\cite{bellorin},\cite{bgmr2}\cite{gmpr}. We impose an additional
constraint to the supermembrane action (\ref{1}), a topological
condition on the map: $\Sigma \to T^{2}$. We denote $X^{r}, r= 1,2$
the maps: $R \times \Sigma \to T^{2}$ and the $X^{m}$ $m=0,\dots,9$
the maps; $R\times \Sigma \to M^{9}$. The topological condition we
impose is the following \cite{torrealba},\cite{mor}: \bea\label{2}
\int_{\Sigma}dX^{r}\wedge dX^{s}=n.
\textrm{Area}_{\Sigma}.\epsilon^{rs}\quad r,s=1,2,\quad n\ne 0 \eea
This global constraint commutes with all first class constraints.
This is so, because any variation $\delta X$ generated by one of the
symmetries of the theory is single-valued on $\Sigma$ and hence \bea
\int_{\Sigma}d(\delta X^{r})\wedge dX^{s}=0, \eea
 since $\delta X^{r}dX^{s}$ is a well defined one-form over $\Sigma$.
Consequently by adding the constraint (\ref{2}) we do not change any
of the original symmetries of the supermembrane. We are only
selecting a sector of the theory. \newline The global restriction
(\ref{2}) ensures that all configurations map a torus $\Sigma$ into
another  torus $T^{2}$, without any degeneration of the mapping. The
left hand member of (\ref{2}) corresponds to the central charge in
the SUSY algebra of the $D=11$ Supermembrane \cite{dwpp}. The
central charge is induced by the irreducible winding condition
\cite{torrealba}. We may also interpret the left hand side of
(\ref{2}) as the integral over $\Sigma$ of the closed two-form
$F=\epsilon_{rs}dX^{r}\wedge dX^{s}$. Condition (\ref{2}) ensures
that $F$ is the curvature of a connection on a nontrivial $U(1)$
principle bundle over $\Sigma$. In fact, the $U(1)$ principle
bundles are classified by $n$.
\newline

We may now fix the Light Cone Gauge, (LCG), since symmetries
have not been altered by the imposition of (\ref{2}).
The gauge fixing procedure is the usual one, we impose
\bea
X^{+}=T^{-2/3}P^{0+}\tau=-T^{-2/3}P^{0}_{-}\tau, \quad P_{-}=P^{0}_{-}\sqrt{W},
\quad \Gamma^{+}\Psi=0
\eea
$\sqrt{W}$ is a time independent density introduced
in order to preserve the density behavior of $P_{-}$ \cite{dwhn}.
 We eliminate $X^{-}, P_{+}$ from the constraints and solve the fermionic
  second class constraints in the usual way. We end up with the lagrangian density:
\bea\label{4} \mathcal{L}=P_{m}\dot{X}^{m}-\mathcal{H} \eea where
the physical hamiltonian in the LCG is given by \bea\label{5}
H=\int_{\Sigma}\mathcal{H}=& \int_{\Sigma}T^{-2/3}\sqrt{W}[\frac{1}{2}(\frac{P_{m}}{\sqrt{W}})^{2}+
\frac{1}{2}(\frac{P_{r}}{\sqrt{W}})^{2}
+\frac{T^{2}}{2}\{X^{r},X^{m}\}^{2}\\ \nonumber &
+\frac{T^{2}}{4}\{X^{r},X^{s}\}^{2}+\frac{T^{2}}{4}\{X^{m},X^{n}\}^{2}]+\textrm{fermionic
terms} \eea subject to the constraints \bea \label{6}
 d(P_{r} dX^{r}+P_{m}dX^{m})=0\eea\bea \label{6a}
& \oint_{\mathcal{C}_{s}}(P_{r} dX^{r}+P_{m}dX^{m})=0 \eea and the
global restriction (\ref{2}). $\mathcal{C}_{s}$ is a canonical
basis of homology on $\Sigma$. The constraints (\ref{6}),(\ref{6a}) are the
generators of area preserving diffeomorphisms homotopic to the
identity. the bracket in (\ref{5}) is given by \bea
\{X^{m},X^{n}\}=\frac{\epsilon^{ab}}{\sqrt{W}}\partial_{a}X^{m}\partial_{b}X^{n},
\eea it is the symplectic bracket constructed from the
non-degenerate two-form \bea \sqrt{W}\epsilon_{ab}d\sigma^{a}\wedge
d\sigma^{b} \eea over $\Sigma$. In 2-dim the area preserving
diffeomorphisms are the same as the symplectomorphisms. There is a
natural election for $W$ on the geometrical picture we have defined.
We consider the $2g$ dimensional space of harmonic one-forms on $\Sigma$. We
denote $dX^{r}$, $r=1,2$, the normalized harmonic one-forms with
respect to $\mathcal{C}_{s}$, $s=1,2$, a canonical basis of homology
on $\Sigma$: \bea \label {7}
\oint_{\mathcal{C}_{s}}d\widehat{X}^{r}=\delta^{r}_{s}. \eea We
define \bea \label{8}
\sqrt{W}=\frac{1}{2}\epsilon_{rs}\partial_{a}\widehat{X}^{r}\partial_{b}\widehat{X}^{s}\epsilon^{ab},
\eea it is a regular density globally defined over $\Sigma$. It is
invariant under a change of the canonical basis of homology.

The action of the new theory, a sector of the original
supermembrane, is now completely defined. We will denote it MIM2.
The interesting property of the MIM2 is that its spectrum is
discrete. Moreover the resolvent of its regularized hamiltonian
operator is compact. This was proven in \cite{bgmr} and also in a
different way using properties of the heat kernel in \cite{br}. The
discretness of the spectrum arises in first place from the absence
of string-like spikes that are present in the original supermembrane
case \cite{dwln}, \cite{dwpp}. This characteristic of the MIM2 is a
consequence of restriction (\ref{2}), and secondly from the linear
dependence of the interacting bosonic-fermionic terms on the bosonic
configuration variables, since it is a supersymmetric action in flat
space.

Another very remarkable fact is that although we still do not know
the  explicit eigenvalues of the complete spectrum of the MIM2,
since it is a theory highly non-linear -in fact it is equivalent to
find the eigenvalues of a symplectic nonperturbative SYM - however
we know that each eigenvalue of the complete spectrum can be bounded
by below up to a perturbation in an operatorial sense
\cite{bgmr}\cite{br}, \cite{bgmr2} by its semiclassical value. The
semiclassical spectrum of the MIM2 was discussed in \cite{stelle}.
It corresponds to the eigenvalues of a supersymmetric harmonic
oscillator whose frequency is given by \bea
\lambda_{\Omega}=\sum_{A\in\Omega} \omega_{A};\quad
\omega_A=\pi^{2}\sqrt{(R^{1}l^{1}a_{2})^{2} +(R^{2}l^{2}a_{1})^{2}}+
susy.\eea where $\Omega=\mathcal{N}\times \mathcal{N}$ is a set of
excited modes  $l^{1},l^{2}$ are the winding numbers and
$a_{1},a_{2}$ are integers characterizing the excitation of the
harmonic oscillator. This result hold for the infinite dimensional
theory and also for a $N$ regularized version of it \cite{gmr}. In
the latest the eigenvalues have in addition a sinusoidal dependence
on the parameters and in the large $N$ limit converge to the exact
ones \cite{bgmr2}. We finally emphasize that the results on this
work are obtained from the complete theory without approximations.

\section{The $SL(2,Z)$ symmetries of the MIM2}
Let us define explicitly the torus $T^{2}$ in the target space.
We consider the lattice $\mathcal{Z}$ on the complex plane $C$,
 \bea \mathcal{Z}: z\to z+2\pi R(l+m\tau)\eea
where $l,m$ are integers, $R$ is a real parameter $ R>0$  and $\tau$
a complex parameter $\tau=Re \tau +i Im\tau,\quad Im\tau>0$, $T^{2}$
is defined by $C/\mathcal{Z}$. $\tau$ is  is the coordinate of the
Teichm\"uller space, for $g=1$ the upper half plane. The
Teichm\"uller space is a covering of the moduli space of Riemann
surfaces. It is in general a
 $2g-1$ complex analytic simply connected manifold. Any flat torus is
 diffeomorphic to $T^{2}$. The conformally equivalent torus are identified
 by $\tau$ modulo the modular group. The parameter $R$ is irrelevant in the
 identification of conformally equivalent classes of compact tori.

Without loosing generality, we may decompose the closed one-forms $dX^{r}$ into
 \bea
dX^{r}=M_{s}^{r}d\widehat{X}^{s}+dA^{r}\quad r=1,2 \eea where
$d\widehat{X}^{s}, s=1,2$ is the basis of harmonic one-forms we have
already introduced, $dA^{r}$ are exact one-forms and $M_{s}^{r}$ are
constant coefficients.  We also define \bea &dX=dX^{1}+idX^{2}\\
\nonumber &dA= dA^{1}+idA^{2} \eea $X^{r}, r=1,2$, are maps:
$\Sigma\to T^{2}$ if and only if they satisfy \bea
\oint_{\mathcal{C}_{s}}dX=2\pi R (l_{s}+m_{s}\tau) \eea $l_{s},
m_{s}, s=1,2$, are integers. This condition is satisfied provided
\bea M_{s}^{1}+iM_{s}^{2}=2\pi R (l_{s}+m_{s}\tau) \eea
Consequently, the most general expression for the maps $X^{r}$,
$ r=1,2$, is \bea\label{9} dX=2\pi R (l_{s}+m_{s}
\tau)d\widehat{X}^{s}+ dA, \eea $l_{s}, m_{s}$, $s=1,2$, arbitrary
integers.
 We now impose condition (\ref{2}).
We obtain after replacing (\ref{9}) into (\ref{2}),
\begin{equation}\label{10}
     n=det\begin{pmatrix} l_{1}& l_{2}\\
                       m_{1} & m_{2}
 \end{pmatrix}
\end{equation}
and \bea \label{11}\textrm{Area}_{\Sigma}=(2\pi R)^{2}Im\tau. \eea
The topological constraint (\ref{2}) only restricts the integral
numbers $l_{s},m_{s}$, $s=1,2$, to satisfy (\ref{10}) with $n\ne 0$.
All configurations (\ref{9}) satisfying condition (\ref{10}) are
then admissible.

In our previous works  we considered $Re \tau=0$ and used the
notation $R_{1}=R$ and $R_{2}=R  Im \tau$ , that is a 2-parameter
torus on the target, besides this point everything is exactly the
same. The MIM2 is invariant under conformal maps homotopic to the
identity (biholomorphic maps). Those are diffeomorphisms which
preserve $d\widehat{X}^{r}, r=1,2$, the harmonic one-forms. $W$ is
then invariant: \bea W^{'}(\sigma)=W(\sigma). \eea Moreover MIM2 is
invariant under diffeomorphisms changing the homology basis, and
consequently the normalized harmonic one-forms, by a modular
transformation on the Teichm\"uller space of the base torus
$\Sigma$. In fact, if \bea\label{12}
d\widehat{X}^{r'}(\sigma)=S^{r}_{s} d\widehat{X}^{r}(\sigma) \eea
then \bea\label{13} \sqrt{W^{'}(\sigma)}d\sigma^{1}\wedge
d\sigma^{2}= \frac{1}{2}\epsilon_{rs}d\widehat{X}^{r'}\wedge
d\widehat{X}^{s'}= \frac{1}{2}\epsilon_{rs}d\widehat{X}^{r}\wedge
d\widehat{X}^{s}= \sqrt{W(\sigma)}d\sigma^{1}\wedge d\sigma^{2} \eea
provided \bea \epsilon_{rs}S_{t}^{r}S_{u}^{s}=\epsilon_{tu} \eea
that is $S\in Sp(2,Z)\equiv SL(2,Z)$. We then conclude that the MIM2
has an additional symmetry with respect to the $D=11$ Supermembrane.
All conformal transformations on $\Sigma$ are symmetries of the MIM2
\cite{br},\cite{bellorin},\cite{bgmmr}, \cite{bgmr2}\cite{gmpr}. We
notice that under (\ref{12}) \bea \label{15}dX\to 2\pi R
(l_{s}+m_{s}\tau)S_{r}^{s}d\widehat{X}^{r}+dA^{'} \eea where
$A^{'}(\sigma^{'})=A(\sigma))$ is the transformation law of a
scalar. That is,
\begin{equation}\label{16}
\begin{pmatrix} l_{1}& l_{2}\\
               m_{1} & m_{2}
 \end{pmatrix}\to \begin{pmatrix} l_{1}& l_{2}\\
                       m_{1} & m_{2}
 \end{pmatrix}\begin{pmatrix} S^{1}_{1}& S^{1}_{2}\\
                       S^{2}_{1} & S^{2}_{2}
 \end{pmatrix}
\end{equation}

$Sp(2,Z)$ acts from the right. Moreover, MIM2 is also invariant
under the following transformation on the target torus $T^{2}$:
\bea \label{17}\tau &\to& \frac{a\tau+b}{c\tau+d}\\
\nonumber  R& \to & R|c \tau +d| \\ \nonumber  A &\to& A
e^{i\varphi_{\tau}}
\\
\nonumber  \begin{pmatrix} l_{1}& l_{2}\\
               m_{1} & m_{2}
 \end{pmatrix}&\to& \begin{pmatrix} a & -b\\
                        -c & d
 \end{pmatrix}\begin{pmatrix} l_{1}& l_{2}\\
               m_{1} & m_{2}
 \end{pmatrix}\eea

where $c\tau+d= |c\tau+d|e^{-i\varphi_{t}}$ and $\begin{pmatrix} a & b\\
                        c & d
 \end{pmatrix}\in Sp(2,Z)$.
 In fact, the potential itself is invariant under (\ref{17}). The
 invariance of the term $\{ X^{r},X^{m}\}^{2}$ may be seen as follows.
We rewrite it as\footnote{c.c denotes complex conjugation.}
 \bea
&\{X^{1},X^{m}\}^{2}+\{X^{2}, X^{m}\}^{2}&=\{2\pi R
(l_{s}+m_{s}\tau)\widehat{X}^{s}+A, X^{m}\}. \textrm{c.c}
 \eea

 We then obtain directly
 \bea
& \{2\pi
R^{'}(l_{s}^{'}+m_{s}^{'}\tau^{'})\widehat{X}^{s} + A^{'}, X^{m}\}.cc=\\
\nonumber &\{ 2\pi R[(l_{s}^{'}d + m_{s}^{'}b)+(l_{s}^{'}c +
m_{s}^{'}a)\tau]\widehat{X}^{s}+ A , X^{m}\}.cc
 \eea
where \bea
\begin{pmatrix} l_{1}^{'} & l_{2}^{'}\\
m_{1}^{'} & m_{2}^{'}
 \end{pmatrix}=\begin{pmatrix} a& -b\\
                       -c & d
 \end{pmatrix}\begin{pmatrix} l_{1}& l_{2}\\
                       m_{1} & m_{2}
 \end{pmatrix}
 \eea
 In the same way it follows that $\{X^{r}, X^{s}\}^{2}$ is invariant
 under (\ref{17}). The hamiltonian density of the MIM2 is then
 invariant under (\ref{17}). The $Sp(2,Z)$ matrix now acts from the
 left of the matrix $\begin{pmatrix} l_{1}& l_{2}\\
                       m_{1} & m_{2}
 \end{pmatrix}$
The two actions from the left and from the right by $Sp(2,Z)$
matrices are not equivalent, they are complementary. The following
remarks are valid.
\subsection{Remark1}
Given  $ \begin{pmatrix} l_{1}& l_{2}\\
                       m_{1} & m_{2}
 \end{pmatrix}$ with determinant  $n\ne 0$, there always exists a
 matrix  $\in SL(2,Z)$ such that its action from the right yields
 \bea
\begin{pmatrix} l_{1}& l_{2}\\
                       m_{1} & m_{2}
 \end{pmatrix}\begin{pmatrix} S^{1}_{1}& S^{1}_{2}\\
                       S^{2}_{1} & S^{2}_{2}
 \end{pmatrix}=\begin{pmatrix} \lambda_{1}& 0 \\
                       \rho & \lambda_{2}
 \end{pmatrix}
 \eea
 where  (of course) $\lambda_{1}\lambda_{2}=n$, and $|\rho|<|\lambda_{2}|$.
   There is an analogous result when the matrix $\in SL(2,Z)$ acts
  from the left. In this case $|\rho|<|\lambda_{1}|$.
  $\rho$ is in general different from zero. One cannot, in general,
  reduce to a diagonal matrix by acting solely from the left or from
  the right.
 \subsection{Remark 2}
 Given $\begin{pmatrix} l_{1}& l_{2}\\
                       m_{1} & m_{2}
 \end{pmatrix}$ with  determinant $n\ne 0$ there always exist matrices
 $\in SL(2,Z)$ such that their action from the left and from the right yields
 \bea
\begin{pmatrix} a & b\\
                        c & d
 \end{pmatrix}
\begin{pmatrix} l_{1}& l_{2}\\
                       m_{1} & m_{2}\end{pmatrix}\begin{pmatrix} S^{1}_{1}& S^{1}_{2}\\
                       S^{2}_{1} & S^{2}_{2}

 \end{pmatrix}=\begin{pmatrix} \lambda_{1}& 0 \\
                       0 & \lambda_{2}
 \end{pmatrix}
 \eea
 where (of course) $\lambda_{1}\lambda_{2}=n$.
 \subsection{Remark 3}
  If $\lambda_{1}$ and $\lambda_{2}$ are not relatively primes,
  the common factor may be absorbed consistently by the parameter $R$.
  We then have the final result. If $\lambda_{1}$ and $\lambda_{2}$
  are relatively primes there always exists matrices $\in SL(2,Z)$
  such that their action from the left and from the right yields
\bea
\begin{pmatrix} a & b\\
                        c & d
 \end{pmatrix}
\begin{pmatrix} \lambda_{1}& 0\\
0 & \lambda_{2}\end{pmatrix}\begin{pmatrix} S^{1}_{1}& S^{1}_{2}\\
                       S^{2}_{1} & S^{2}_{2}

 \end{pmatrix}=\begin{pmatrix} n& 0 \\
                       0 & 1
 \end{pmatrix}
 \eea
This proves that although we may have arbitrary winding numbers
 $\begin{pmatrix} l_{1} & l_{2}\\
                        m_{1} & m_{2}
 \end{pmatrix}$ the symmetries of the MIM2 allow to reduce
 everything to the central charge integer $n$ (\ref{2}). The MIM2
 satisfies then the requirement raised in \cite{schwarz} , footnote
 6. Consequently, there is a canonical formula for the harmonic sector
 of the maps. The general expression for the $dX$ maps is then
 \bea
dX=2\pi R (n d\widehat{X}^{1}+ \tau d\widehat{X}^{2})+dA
 \eea
 or in components
 \bea\label{18}
&dX^{1}=2\pi R (n dX^{1}+ Re \tau d\widehat{X}^{2})+dA^{1}\\
\nonumber & dX^{2}=2\pi R ( Im\tau d\widehat{X}^{2})+ dA^{2}
 \eea

 There is also a formulation in which $n$ is attached to
 $d\widehat{X}^{2}$ instead of $d\widehat{X}^{1}$.

We may now determine the subgroup of (\ref{17}) which together with the $SL(2,Z)$
transformations on the base manifold, yield a covariant transformation law for
the harmonic part of $dX$,
\bea\label{nt11}
dX_{h}=2\pi R (n d\widehat{X}^{1}+\tau d\widehat{X}^{2}).
\eea
It turns out to be the subgroup  $\begin{pmatrix} a & b\\
                        c & d
 \end{pmatrix}\in SL(2,Z)$ where $b=nb_{1}$ and $b_{1}$ an integral number.
A fundamental region $F$ on the upper half plane associated to those
transformations is obtained from the canonical fundamental region
associated to the $SL(2,Z)$ by taking discrete translations on the
real direction of values $\pm 1,\pm 2, \dots,\pm (n-1)$. $F$ is
contained in the union set of the fundamental region and its
translations. This subgroup of transformations
preserve the form (\ref{nt11}) and leaves invariant the hamiltonian.\\
The harmonic map (\ref{nt11}) is a minimal immersion from $\Sigma$ to $T^{2}$ on the target,
moreover it is directly related to a holomorphic immersion of $\Sigma$ onto $T^{2}$.
The extension of the theory of supermembranes restricted by the topological constraint
to more general compact sectors in the target space is directly related to the existence of
those holomorphic immersions, because of this property we denote this sector of the $D=11$
supermembrane MIM2.

 We have shown that the MIM2 has two $SL(2,Z)$ symmetries.
These symmetries explain the necessary identification mentioned
in \cite{schwarz}, footnote 6.
 One is
 associated to the conformal invariance on the base manifold. It
 allows to identify the winding modes. The other $SL(2,Z)$ symmetry (\ref{17})
 acts on the Teichm\"uler coordinate $\tau$. However since the other
 parameter $R$ is also involved in the transformation, the
 equivalence classes of tori under this transformation are not the
 conformally equivalent classes. We will show in the following
 sections, using these two $SL(2,Z)$ symmetries, that the mass
 contribution of the string states in the MIM2 exactly agree with
 the perturbative mass spectrum of $(p,q)$ $IIB$ and $IIA$
 superstring.

\section{The irreducible winding and KK contribution to the mass$^2$}
We start writing explicitly all the contributions to the $mass^{2}$
arising from the MIM2 theory. We consider
$mass^{2}=-P_{\mu}P^{\mu}$. The first term to evaluate is the pure
harmonic contribution in (\ref{18}) to the $mass^{2}$. It arises
from the term in (\ref{5}) \bea
T^{-2/3}\int_{\Sigma}\sqrt{W}\frac{T^{2}}{4}\{X^{r}_{h},X^{s}_{h}\}^{2}\eea
where $X^{r}_{h}$ denotes the harmonic part in (\ref{18}). The
factor $T^{-2/3}$ cancels an inverse factor from the $2P_{+}P_{-}$
term. The result is \bea\label{19} mass^{2}=T^{2}((2\pi R)^{2}n
Im\tau)^{2}+\dots \eea The dots represent additional terms which we
will shortly write explicitly. This contribution is usually obtained
by considering the tension $T$ as the $mass/area$, times the total
area $nA$, $A=(2\pi R)^{2} Im \tau$. This is correct provided the
contribution of all winding modes resumes in the factor $n$, which
we have shown explicitly to be the case for the MIM2.

The harmonic sector is related by the condition (\ref{2}) to a
nontrivial $U(1)$ principal bundle, classified by the winding number
$n$. This is the contribution of these nontrivial complex bundles to
the $mass^{2}$. We will interpret the KK quantization condition in a
similar way.

Suppose we are compactifying on  $S^{1}\times S^{1}$ and the
associated momenta are $p_{r}$ $r=1,2$. Then for $r=1,2$ we have, we
do not use the index explicitly \bea\label{20} P=\int_{\Sigma} p
d\sigma^{1}\wedge d\sigma^{2}. \eea  where $p$ is a scalar density.
It may always be expressed as the dual to a two-form $F$,
\bea\label{22} Rp=\epsilon^{ab}F_{ab} \eea
 $R$ is the radius of the circle. (\ref{20}) may be rewritten
 \bea\label{21}
RP=\int_{\Sigma} F , \eea we take $P$ to have dimensions of mass,
and $F$ to be dimensionless. The quantization condition is to set
\bea\label{21} \int_{\Sigma}F=2\pi m.\eea This is exactly the Weil
condition on a closed two form, in dimension 2 $F$ is closed, to
ensure the existence of a $U(1)$ principle bundle and a connection
whose curvature is $F$. We then have for each mode\bea\label{22}
R\frac{p}{\sqrt{W}}=\frac{\epsilon^{ab}F_{ab}}{\sqrt{W}}=F^{*}=m,
\eea and the associate conjugate coordinate $T^{-2/3}p=\dot{X}$,
\bea X=T^{-2/3}\frac{m}{R}\sqrt{W}t. \eea That is, there is a
geometrical global contribution from those configurations. In the
case of the MIM2 we first have to change to a frame where the maps
are onto circles. We have from (\ref{18}) \bea \frac{1}{2\pi
R}\oint_{\mathcal{C}_{s}} \mathbb{M}^{-1} \begin{pmatrix} dX^{1}\\
                       dX^{2}\end{pmatrix}=\oint_{\mathcal{C}_{s}}
                        \begin{pmatrix} n d\widehat{X}^{1}\\
                       d\widehat{X}^{2}\end{pmatrix}\eea

where
$\mathbb{M}= \begin{pmatrix} 1& Re\tau\\
                       0 & Im\tau\end{pmatrix}$.
The associated maps are then onto circles. The corresponding momenta
are $p_{s}M_{r}^{s}$. The quantization condition is then \bea
R\int_{\Sigma}p_{s}M_{r}^{s}d\sigma^{1}\wedge d\sigma^{2}
=2\pi m_{r}\quad r=1,2\eea
 $m_{r}$ integers. We then have $p_{s}=p_{s}^{0}\sqrt{w}$ where
 \bea
p_{s}^{0}=\frac{m_{r}}{R}(M^{-1})^{r}_{s},
 \eea
that is \bea\label{23} p_{1}^{0}&=&\frac{m_{1}}{R}\\ \nonumber
p_{2}^{0}&=&-\frac{m_{1}Re \tau +m_{2}}{R Im \tau}\eea
 We now observe that the only contribution to the total momenta in
 those directions arises solely from the nontrivial line bundles.
 There is no contribution from the local degrees of freedom. In fact we may
 always impose the gauge condition
 \bea\label{24}
\int_{\Sigma}A^{r}\sqrt{W}d\sigma^{1}\wedge d\sigma^{2}=0
 \eea
In fact \bea \delta X^{r}=\{\xi, X^{r}_{h}\}+\{\xi, A^{r}\}, \eea
and \bea \int_{\Sigma}\{\xi, A^{r}\}\sqrt{W}d\sigma^{1}\wedge
d\sigma^{2}=0 \eea  since $A^{r}$
 is single-valued on $\Sigma$. However
 \bea\label{25}
\int_{\Sigma}\{\xi_{h}, X^{r}_{h}\}\sqrt{W}d\sigma^{1}\wedge
d\sigma^{2}\ne 0
 \eea
 provides a non zero contribution. Moreover since
 $\xi_{h}=a_{r}(t)\widehat{X}^{r}$, the two parameters $a_{r}(t)$
 allow to impose the gauge fixing condition (\ref{24}). Hence the
 contribution of $\dot{A}^{r}$ to the overall momenta is zero. In (\ref{25})
  the parameter $\xi_{h}$ is associated to the generator
  (\ref{6a}), $d\xi_{h}$ is a harmonic one-form. the
  bracket in (\ref{25}) is constructed from $d\xi$. We then have
  \bea\label{26}
p_{r}=p_{r}^{0}\sqrt{w}+\Pi_{r},
  \eea
where \bea\label{27} \int_{\Sigma}\Pi_{r}d\sigma^{1}\wedge
d\sigma^{2}=0. \eea The contribution of the KK modes to the
$mass^{2}$ arises from the substitution of (\ref{26}) into the
$p_{r}^{2}$ term in (\ref{5}), and the use of (\ref{27}). We obtain
\bea\label{28} mass^{2}=T^{2}((2\pi R)^{2}n Im \tau)^{2}+
\frac{1}{R^{2}}((m_{1}^{2}+(\frac{m_{2}-m_{1}Re\tau}{Im
\tau})^{2}))+ T^{2/3}H\eea where the $H$ is the hamiltonian of the
MIM2. It was first obtained and analyzed as a symplectic non
commutative theory in \cite{mor} and proved to have
discrete spectrum in \cite{bgmr},\cite{br}. It has the same
expression (\ref{5}) without of course the global modes that
already have being separated. We will discuss $H$ in the next section.
The two explicit terms on the right hand side of (\ref{28}) are
exactly the ones in \cite{schwarz} which allow the identification of
$\tau$ to the moduli of the $(p,q)$ class of $IIB$ superstring
solutions. In the next section we will obtain the complete
identification of the mass formulas beyond these two terms.

%%%%%%%%%%%%%%%%%%%%%%
The KK term in (\ref{28}) can be written as \bea (\frac{\vert
m_{1}\tau -m_{2}\vert}{R Im \tau})^{2}=(\frac{m\vert
q_{1}\tau-q_{2}\vert}{R Im \tau})^{2} \eea

Where $q_1$ and $q_2$ are relatively prime integral numbers.
Under (\ref{17}) \bea &\tau \to \tau^{'}=\frac{a\tau +b}{c\tau +d}\\
\nonumber & R\to R^{'}=R\vert c\tau +d\vert\eea

and we add now, \bea (q_{1},-q_{2})\to (q_{1}^{'},-q^{'}_{2})\begin{pmatrix} d& -b\\
                       -c & a\end{pmatrix} \eea
We then have

\bea \frac{\vert q_{1}^{'}\tau^{'}-q_{2}^{'}\vert}{R^{'}Im
\tau^{'}}=\frac{\vert q_{1}\tau-q_{2}\vert}{R Im \tau }, \eea

we have thus completed the transformation law (\ref{17}).

\section{String states in the MIM2 theory}

We will consider in this section the states associated to
configurations which depend on one local coordinate. Instead of
considering local coordinates $\sigma^{a}$ $a=1,2$, we will work
with $\widehat{X}^{r}$ $r=1,2$. The Jacobian of the transformation
is
$\frac{1}{2}\epsilon^{ab}\partial_{a}\widehat{X}^{r}\partial_{b}\widehat{X}^{s}\epsilon_{rs}$
 which is nonsingular over $\Sigma$. The local coordinates
 \bea
\sigma^{1},\sigma^{2} \to \widehat{X}^{1}, \widehat{X}^{2}
 \eea
on single valued scalar fields is then well defined. We then
consider within the physical configurations of the MIM2, the
string configurations \bea X^{m}=X^{m}(\tau,
q_{1}\widehat{X}^{1}+q_{2}\widehat{X}^{2}),\quad  A^{r}=A^{r}(\tau,
q_{1}\widehat{X}^{1}+q_{2}\widehat{X}^{2}) \eea where $q_{1},q_{2}$
are relative prime integral numbers. $X^{m}, A^{r}$ are scalar
fields on the torus $\Sigma$, a compact Riemann surface, hence they
may always be expanded on a Fourier basis in term of a double
periodic variable of that form. The restriction of $q_{1}, q_{2}$ to
be relatively prime integral numbers arises from the global
periodicity condition. On that configurations all the brackets \bea
\{X^{m},X^{n}\}=\{X^{m}, A^{r}\}=\{A^{r},A^{s}\}=0 \eea vanish. The
hamiltonian of the string configurations becomes \bea H \vert _{SC}= T^{-2/3}\int_{\sqrt{W}}[
\frac{1}{2}(\frac{P_{m}}{\sqrt{W}})^{2}+\frac{1}{2}(\frac{\Pi_{r}}{\sqrt{W}})^{2}+
\frac{T^{2}}{2}\{X^{r}_{h},X^{m}\}^{2}+\frac{T^{2}}{2}\{X^{r}_{h},A^{s}\}^{2}]
\eea

where $X^r_h$, $r=1,2$, denote the harmonic sector of the one-form
$dX^r$, $r=1,2$, subject to the constraints \bea
_{}\{X^{r}_{h},\frac{\Pi_{r}}{\sqrt{W}}\}=0 \qquad
\oint_{\mathcal{C}_{s}}(\frac{P_{r}dX^{r}}{\sqrt{W}}+
\frac{P_{M}dX^{M}}{\sqrt{W}})=0 \eea

where $\mathcal{C}_{s}$ is the basis of homology associated to the
harmonic basis $\widehat{X}^{r}$. The local constraint may be
explicitly solved: \bea \frac{\Pi_{r}}{\sqrt{W}}=T^{2/3}\{
X^{s}_{h}, \frac{\Pi}{\sqrt{W}}\}\epsilon_{rs} \eea  in terms of an
unconstrained field $\Pi$. We then have for the kinetic term \bea
&\int_{\Sigma} p_{r}\partial_{t}{X}^{r}=\int_{\Sigma}
\Pi_{r}\partial_{t}{A}_{r}+\int_{\Sigma }\partial_{t}{(\sqrt{W}p_{r}^{0}A^{r})}=\\ \nonumber &
\int_{\Sigma} (T^{-2/3}\dot{\Pi})(T\{X_{h}^{s},
A^{r}\}\epsilon_{rs})+\int_{\Sigma}
\partial_{t}{(\sqrt{w}p^{0}_{r}A^{r})}.\eea

The total time derivative may be eliminated from the Hamiltonian
formulation.  This property is valid in general for the MIM2
hamiltonian. We have then solved the local constraint and
canonically reduced the theory to a locally unconstrained theory
where the new conjugate pair is: \bea (T^{-2/3}\Pi),\quad
T\{X^{s}_{h},A^{r}\}\epsilon_{rs}.\eea

We then have \bea
H|_{SC}=T^{-2/3}\int_{\Sigma}\sqrt{W}[\frac{1}{2}(\frac{P_{M}}{\sqrt{W}}
)^{2} + \frac{T^{2}}{2}\{X^{r}_{h},X^{m}\}^{2}]+ \textrm{fermionic
terms} \eea where now $M=1,\dots,8$. It is a SUSY harmonic
oscillator. It has the same number of bosonic and fermionic creation
and annihilation operators, with cancelation of the zero point
energy contribution \cite{stelle}. The potential has the explicit form
\bea\label{71} (2\pi R)\frac{T^{2}}{2}\vert \{ n\widehat{X}^{1}+\tau
\widehat{X}^{2}, X^{M}\}\vert^{2} \eea

We now perform a change on the canonical basis of homology and a
corresponding change of the basis of harmonic one-forms: \bea\label{72}
&d\widetilde{X}^{1}=q_{1}d\widehat{X}^{1}+q_{2}d\widehat{X}^{2}\\
\nonumber &
d\widetilde{X}^{2}=nq_{3}d\widehat{X}^{1}+q_{4}d\widehat{X}^{2},
 \eea
given $q_1$ relatively prime to $q_2$ and to n, there always exist
$q_3$ and $q_4$ such that expression $\begin{pmatrix} q_{1}& q_{2}\\
                       n q_{3} & q_{4}\end{pmatrix}\in SL(2,Z)$.

If $q_{1}$ and $n$ have a common factor we attached it to
$\widehat{X}^{1}$ in (\ref{71}),(\ref{72}) and proceed with the remaining
factors $\widetilde{q}$ and $\widetilde{n}$.

The harmonic sector when expressed in the new harmonic basis
satisfies \bea 2\pi R T \vert n d\widehat{X}^{1}+ \tau
d\widehat{X}^{2}\vert =2 \pi \widetilde{R} T \vert  n
d\widetilde{X}^{1}+\widetilde{\tau} d\widetilde{X}^{2}\vert\eea
where as in (\ref{17}) \bea\label{74}\widetilde{R}= R\vert q_{4}-\tau q_{3}
\vert,\qquad \widetilde{\tau}=\frac{-n q_{2}+\tau q_{1}}{q_{4}-\tau
q_{3}}. \eea It is then in its canonical form. In the case when $n$
and $q_1$ have a common factor the formulas are the same with $n$
and $q_1$ replaced by $\widetilde{n}$ and $\widetilde{q_{1}}$. We
now consider the global constraint. We take the homology basis to be
the corresponding one to the new harmonic basis. We are allowed to
do that since the global constraint has to be imposed on any basis
of homology. The constraint associated to the element of the basis
$\mathcal{C}_{2}$ does not have contributions from the local degrees
of freedom since they depend only on $\widetilde{X}^{1}$. We then
obtain $m_2=0$. The constraint associated to expression
$\mathcal{C}_{1}$ yields \bea m_{1}n =N_{R}- N_{L} \eea
 the
level-matching condition and \bea (P^{0}_{1})^{2}+ (P^{0}_{2})^{2}=
(\frac{\vert \widetilde{\tau}\vert}{\widetilde{R} Im
\widetilde{\tau}})^{2}. \eea There are also non-stringy states
contributing to the matching level conditions. We are freezing those
contributions and considering only the string contributions. Let us
go on the final step of the argument. We started in (\ref{71}) with
$\tau$ in the fundamental region $F$  described at the end of
section 3. $\widetilde{\tau}$ in (\ref{74}) is exactly a general
transformation with the subgroup of $SL(2,Z)$ associated to $F$.
$\widetilde{\tau}$ is then a generic point on the upper half plane.
But any point on the upper half plane may be obtain from the
fundamental region of $SL(2,Z)$ by a M{\"o}bius transformation \bea
\widetilde{\tau}=\frac{q\tau -p}{Q\tau +P}
 \eea
 where
 $\begin{pmatrix} q& -p\\
                       Q & P\end{pmatrix}\in SL(2,Z)$.
(the "-" is only conventional). We now notice that the reduced
expression of the hamiltonian is covariant under that more general
transformation. We then obtain the final expression for the mass
contribution of the string states:
 \bea\label{78} M_{11}^{2}\vert_{SC}= (n
T_{11}(2\pi R_{11} Im\tau)^{2}+(\frac{m\vert q\tau -p\vert}{R_{11}
Im \tau
}))^{2}+ 8\pi^{2} R_{11}T_{11}\vert q\tau -p \vert
(N_{L}+N_{R}) \eea where $(p,q)$ are relatively prime. We have
denoted $R_{11}, \tau$ the transformed variables, and $T_{11}=T$ the
eleven dimensional tension. We notice that $(p,q)$ may be interpreted as
the wrapping of the membrane around the two cycles of the target torus.
In fact, we can always re-express the KK-term in terms of
$\widetilde{\tau}=\frac{q\tau-p}{Q\tau+P}$, as
\bea
\frac{m\vert\widetilde{\tau}\vert}{\widetilde{R}Im\widetilde{\tau}}
\eea
using(\ref{17}). The corresponding change in the harmonic sector is
\bea
dX_{h}=(qmd\widetilde{X}^{1}+pd\widetilde{X}^{2})
+\widetilde{\tau}(-Qnd\widetilde{X}^{1}+Pd\widetilde{X}^{2}),
\eea
the hamiltonian is invariant under that change. $p,q$ and $Q,P$ are now the
winding numbers of the supermembrane. Given $p,q$ there always exist
$Q$ and $P$
with the above property, although the correspondance is not unique. The $(p,q)$
type IIB strings may indeed be interpreted as different wrappings of the MIM2.
This nice interpretation was first given in \cite{schwarz} from a heuristic point of view.

  We may now compare the mass formula for
the string states in the MIM2  with respect to the mass formula of
the $(p,q)$ $IIB$ and $IIA$ strings.\\

The $(p,q)$ $IIB$ string compactified on a circle of radius $R_{B}$
has tension\cite{schwarz} \bea T_{(p,q)}^{2}=\frac{\vert
q\lambda_{0}-p  \vert^{2}}{Im\lambda_{0}}T^{2} \eea where
$\lambda_{0}=\xi_{0}+ie^{-i\phi_{0}}$. $\xi$ and $\phi$ can be
identified to the scalar fields of the type IIB theory, $\phi$
corresponds to the dilaton fields. $\lambda_{0}$ is the asymptotic
value of $\lambda$ -the axion-dilaton of the type IIB theory-
specifying the vacuum of the theory. The perturbative spectrum of
the $(p,q)$
 IIB string is \cite{schwarz},
\bea\label{82} M_{B}^{2}=(\frac{n}{R_{B}})^{2}+(2\pi R_{B}m
T)^{2}+T_{(p,q)}4\pi (N_{L}+N_{R}) \eea If we use following
\cite{schwarz} a factor $\beta^{2}$ to identify term by term
 both mass formulas, since there were obtained using different metrics, one gets
\bea\label{83} \tau=& \lambda_{0}\\ \nonumber \beta^{2}=&
 \frac{T_{11}A_{11}^{1/2}}{T}\\ \nonumber R_{B}^{-2}=& T T_{11}
A_{11}^{3/2} \eea where $A_{11}=(2\pi R_{11})^{2}Im\tau$ is the area
of the torus. These relations were obtained in \cite{schwarz} by
comparing the first two terms on the right hand side of (\ref{78})
and (\ref{82}). They were obtained by counting modes under some
assumptions on the supermembrane wrapping modes, as mentioned on one
of the footnotes \cite{schwarz}. Here we have derived the
expressions from a consistent definition of the MIM2. In addition
the comparison of the third terms in (\ref{78}) and (\ref{82}) is
completely consistent with relations (\ref{83}). \newline We notice
that $A_{11},\beta$ and $R_{B}$  are invariant under (\ref{17}),
$T_{p,q}$ is also invariant provided $(p,q)$ transforms under an
associated $SL(2,Z)$ matrix.

The identification of (\ref{78}) to the mass formula of IIA string
compactified on a circle of radius $R_{A}$ and tension $T_{A}$ may
also be performed. In order to have a consistent identification one
has to take $Re\tau=0$, $p=1$ and hence $q=0$ in (\ref{78}). The
mass formula for the perturbative spectrum of type IIA is \bea
M_{A}^{2}=(\frac{m}{R_{A}})^{2}+(2\pi R_{A}n T_{A})^{2}+T_{A}4\pi
(N_{L}+N_{R}) \eea Identification after the limit process explained
in section 6 of the winding contributions and KK ones
 using a factor $(\beta\gamma)$ to compare the $mass^{2}$ formulas,
since they are obtained using different metrics, yields
\bea
& R_{A}=\beta\gamma R_{11}\\ \nonumber &
T_{A}=\gamma^{-2}(Im\tau)^{1/2} T
\eea
which imply
\bea
(2\pi R_{A}R_{B})=(\frac{1}{T_{A}T Im \tau^{1/2}})^{1/2}
\eea
We have thus obtained the $(p,q)$ IIB and IIA perturbative spectrum,
when compactified on circles $R_{B}$ and $R_{A}$ respectively, from the string states on
the MIM2. \newline
%%%%%%%%%%%%
\subsection{$SL(2,Z)$ (p,q)-strings on IIA}
We introduce now the dual tori $\widetilde{T}^{2}$. $T^{2}$ was
defined in terms of the moduli $(R, \tau)$. The moduli of
$\widetilde{T}^{2}$, $(\widetilde{R}, \widetilde{\tau})$ are defined
by the relation
\bea
z\widetilde{z}=1
\eea
where the dimensionless variables $z, \widetilde{z}$ are
\bea
 z=T_{11}A_{11}\widetilde{Y}\\ \quad
\widetilde{z}=T_{11}\widetilde{A}_{11}Y
\eea

$A_{11}=(2\pi R)^{2} Im \tau$, $\widetilde{A}_{11}=(2\pi
\widetilde{R}^{2} Im \widetilde{\tau})$ are the area of the $T^{2}$
and $\widetilde{T}^{2}$ and \bea Y=\frac{R Im
\tau}{\vert\tau\vert}\quad \widetilde{Y}=\frac{ \widetilde{R}
Im\widetilde{\tau}} {\vert\widetilde{\tau}\vert}, \eea We now look
for the stringy states of the MIM2 wrapping on a $\widetilde{T}^{2}$
torus. Their contribution to the mass formula is given by \bea
\widetilde{M}_{11}^{2}=(m
T(2\pi\widetilde{R})^{2}Im\widetilde{\tau})^{2}+
(\frac{n\vert\widetilde{\tau}\vert}{\widetilde{R}Im\widetilde{\tau}})^{2}+
2\pi\widetilde{R}T\vert\widetilde{\tau}\vert 4\pi (N_{L}+N_{R}) \eea
We may now match the winding term of $M_{11}^{2}$ with the KK term
of the
 $\widetilde{M}_{11}^{2}$ and
viceversa. We obtain only one relation, the
 other two are identically satisfied,
which gives the proportionality constant $\alpha$ between $M_{11}$
and $\widetilde{M}_{11}$:
\bea
\alpha^{2}=\frac{\widetilde{z}}{z}
\eea

It defines a T-duality relation on MIM2.
We may take $\widetilde{\tau}\to \frac{a\tau+b}{c\tau+d}$, $R\to
R\vert c\tau+ d\vert$ and obtain the corresponding
$\widetilde{M}_{11}$ formula.
 For the $(p.q)$ string states contribution to $M_{11}^{2}$ there is associated a multiplet
 of infinite states on the
 MIM2 wrapping on $\widetilde{T}^{2}$, which should correspond to D-brane bound
states in lower dimensions. The precise relation has not yet been determined.
There is an interesting nonlinear relation between the $SL(2,Z)$ on $T^{2}$ and
 the $SL(2,Z)$ on $\widetilde{T}^{2}$. We will report on it elsewhere.

%%%%%%%%%%%%%%%%%%%%%%%
\subsection{The decompactification limits}
We consider first the limit when $R_{B}\to \infty$, $\tau$ in the upper half plane
and $n\to \infty$ as defined below. If $R_{B}\to \infty$ then $A_{11}=(2\pi R_{11})^{2} Im \tau \to 0$ and
necessarily $R_{11}\to 0$. The KK term in the $M_{11}^{2}$ of the MIM2 behaves as\bea
\frac{m^{2}\vert q\tau -p\vert^{2}}{(R_{11} Im \tau)^{2}}\sim \frac{m^{2}}{R_{11}^{2}}\to\infty.
\eea
 The winding term in $M_{11}^{2}$ behaves as
\bea (nA_{11}T_{11})^{2}\sim n^{2}R_{11}^{4}\sim
(\frac{1}{R_{11}})^{2}\to\infty\quad \textrm{if}\quad n\sim
\frac{1}{R_{11}^{3}}, \eea we thus consider $n\to\infty$ in the
above way. The hamiltonian contribution, first in the semiclassical
approximation, to the $M_{11}^{2}$ in terms of $p_{1},p_{2}$ the two
integers identifying the MIM2 modes, is \bea\label{5.11}
\sum_{p_{1},p_{2}} 2\pi R_{11}T_{11}\vert np_{2}+\tau p_{1}\vert
\eea where $R_{11}Im \tau\to 0$  while $R_{11} n\sim
\frac{1}{R_{11}^{2}}\to\infty$. The hamiltonian contribution of the
exact theory is bounded from below, up to perturbative (in the
operatorial sense) terms by its semiclassical hamiltonian
\cite{bgmr},\cite{br},\cite{bgmr2}. That means that eigenvalues of
the exact theory are bounded from below by the semiclassical ones,
up to small perturbations. These perturbative terms arise from the
fermionic cubic interactions and were rigourously characterized in
\cite{bgmr},\cite{br}.\newline We then conclude that the
contribution to the energy of (\ref{5.11}) is mainly from the
$p_{1}$ modes only. That is, given a fixed value of the energy $E$,
for small enough values of $R_{11}$ the eigenvalues less than $E$
arise from the $p_{1}$ modes only. The total energy is then \bea
2\pi R_{11}T_{11}\vert \tau \vert 4\pi (N_{L}+N_{R}). \eea It may be
expressed no in terms of $\widetilde{\tau}$ on the fundamental
region and
 from (\ref{17}) we obtain
\bea 2\pi\widetilde{R}_{11}T_{11}\vert q\widetilde{\tau}-p\vert 4\pi
(N_{L}+N_R) \eea as expected for the energy spectrum of 10
dimensional $(p,q)$ IIB strings, in the eleven dimensional metric.
\newline
Let us now consider the limit when $R_{A}\to\infty$ and
consequently $A_{11}\to\infty$. The limit is defined by taking
$Im\tau\to\infty$, $Re\tau\to\infty$, for any finite $n$ and
$R_{11}\to 0$. We then have \bea \frac{m^{2}\vert q\tau
-p\vert^{2}}{(R_{11}Im\tau)^{2}}&\sim&\frac{m^{2}}{R_{11}^{2}}\to\infty\\
\nonumber (nA_{11}T_{11})^{2}&\sim &
R_{11}^{4}Im\tau^{2}\sim\frac{1}{R_{11}}\to\infty \eea provided
$Im\tau\sim\frac{1}{R_{11}^{3}}$, which we assume. The hamiltonian
contribution to $M_{11}^{2}$ is then \bea &2\pi
R_{11}T_{11}Im\tau\sim \frac{1}{R_{11}^{2}}\to\infty\\ \nonumber &
2\pi R_{11}T_{11}n\to 0 \eea By the same reason as before the
contribution for the small $R_{11}$ is mainly from the
 $p_{2}$ modes only, with a tension independent of $\tau$.
In both decompactified limits we have considered the right hand
member of the topological constraint becomes \bea
nA_{11}\sim\frac{1}{R_{11}}\to\infty, \eea they correspond to limit
cases. We will comment on these limits in the conclusions.
\subsection{The wrapping of the MIM2 on a circle}
Let us consider now the wrapping of the MIM2 on a circle but with a finite value
 of the topological constraint. We will show that this is possible if the MIM2 wraps only once.
It then answers, in the context of the MIM2, the puzzle raised in \cite{schwarz}:
 ``Why the membrane can be wrapped on a torus any number of times but only one on a circle?''
The limit we are going to consider was first discussed in
\cite{stelle}. It considers the limit when the radius of one of the
circles on the target space goes to zero but the supermembrane wraps
infinite number of times around it in a way that $l_{1}R_{11}=R_{1}$
finite and $R_{11}Im \tau=R_{2}$ finite. It thus imply $Im\tau \to
\infty$ . We then have for the right hand member of the topological
constraint \bea nA_{11}=l_{2}A \eea where $A=(4\pi^{2}R_{1}R_{2})$
and $l_{2}$ is the winding number around the circle of radius
$R_{2}$. To analyze this limit it is better to consider an
expression of the hamiltonian in terms of the windings $l_{1}$ and
$l_{2}$: $l_{1}l_{2}=n$ explicitly. According to the remarks in
section 3 we can always work in terms of $n$ or equivalently in
terms of $l_{1}$ and $l_{2}$. All possible decompositions of $n$ in
terms of $l_{1}$ and $l_{2}$ are equivalent as a consequence of the
symmetries of the MIM2. We have now for the winding contribution,
\bea (nA_{11}T_{11})^{2}=(l_{2}A T_{11})^{2}, \eea and for the KK
contribution, it is finite only when $q=0,p=1$: \bea
(\frac{m}{R_{2}})^{2}. \eea That is, the only possibility is to have
$q=0, p=1$. Finally the hamiltonian contribution at the
semiclassical level is \bea \sum_{p_{1},p_{2}}2\pi R_{2}T_{11}\vert
i\frac{R_{1}}{R_{2}}p_{2}+l_{2}p_{1} \vert, \eea which in the string
limit we consider below becomes exact. We may now interpret
$\frac{R_{2}}{R_{1}}$ as $Im\tau$ and obtain the canonical
contribution of a MIM2 with $Re\tau=0$ and winding number $l_{2}$.
We consider now together with $R_{11}\to 0$ no dependence on
$\widehat{X}^{1}$, that is on the $p_{1}$ modes. We are considering
the stringy states obtained from the above MIM2 by freezing the
remaining membrane states. We then obtain exactly the nine
dimensional $mass^{2}$ contribution of the IIA superstring
compactified on a circle of radius $R_{2}$ with winding $l_{2}$, in
the 11 dimensional metric. In fact, the matching level condition
arising from the MIM2 constraints in the
 above limit is
\bea
ml_{2}=N_{R}-N_{L}.
\eea
The wrapping of the supermembrane on a circle to obtain the type IIA theory is
 then only possible if $p=1,q=0$.

\section{Discussion and conclusions}
We obtained the quantum symmetries of the MIM2 theory, a sector of
the $D=11$ supermembrane. There is a $SL(2,Z)$ symmetry realized by
the area preserving diffeomorphisms not homotopic to the identity.
It acts as the modular group on the Teichm{\"u}ller space associated
to the base manifold. This symmetry transforms equivalent classes of
maps, under area preserving diffeomorphisms homotopic to the
identity (gauge transformations), into equivalent classes. The
hamiltonian is invariant under these transformations. In addition
the hamiltonian density is invariant under a $SL(2,Z)$
transformations acting on the moduli of the target space.  Although
this transformation on the Teichm{\"u}ller parameter $\tau$ of the
target torus is a M{\"o}bius transformation, the equivalence classes
of tori under it are not the conformal classes. Both transformations
are not equivalent and both are relevant in the analysis of the MIM2
hamiltonian. Consequently the usual statement that the $SL(2,Z)$
duality symmetry of $IIB$ superstrings is associated to the
reparametrization invariance on the M-theory is not precise. We
explicitly showed that the duality symmetry of $IIB$ superstrings
has its origin in the supermembrane theory and it is related, at
least in the MIM2 sectors to the space of holomorphic immersions, or
more generally minimal immersions, from the base manifold to the
target space. The space of holomorphic immersions is a very
interesting one and was discussed in a different context in
\cite{witten3}. The holomorphic immersion may be constructed, in the
case discussed in this work, in terms of the harmonic maps. The
above symmetries allow to express the harmonic maps in a canonical
form. The corresponding harmonic one-forms are covariant under a
subgroup of $SL(2,Z)$: the matrices  $\begin{pmatrix} a& b\\
                       c & d\end{pmatrix}\in SL(2,Z)$ with
                       $b=nb_{1}$, where $b_{1}$ is an integral
number and $n$ the winding number of the MIM2. $n$ is introduced in
the theory through the topological constraint. The corresponding
fundamental region on the upper half plane may be determined from
the fundamental region of the modular group. In particular, the
symmetries of the MIM2 allow to understand why the different winding
modes of the supermembrane are equivalent. This point was left as an
assumption in \cite{schwarz}. From the action of the MIM2 we
obtained directly the winding and KK contributions to the nine
dimensional $mass^{2}$ formula, in exact agreement with
\cite{schwarz}. We obtain all $(p,q)$ type $IIB$ string states
within the MIM2. This is done by considering the matching level
constraints of the MIM2 and the string configurations on it. By
freezing the pure membrane states satisfying the matching level
condition, the contribution of the stringy configurations to the
$mass^{2}$ may be determined. This can be done from the full theory
without any approximation, nor any particular limit, using only its
symmetries. This $mass^{2}$ contribution is exactly the $mass^{2}$
formula for the $(p,q)$ type $IIB$ superstring compactified on a
circle, including the winding, KK and oscillator contributions. We
also obtained the perturbative mass spectrum of $IIA$ superstring
compactified on a circle from the MIM2. This was done following
\cite{stelle} by taking the limit $R_{11}\to 0$, $Im \tau \to
\infty$, in such a way that the topological constraint has a finite
righthand member. This limit is only consistent if $p=1,q=0$,
showing that the supermembrane can wrap a torus any number of times
but  a circle only once. This was a puzzle raised in \cite{schwarz}.
The solution in \cite{schw2} has the same conclusion although the
limits taken there are different from ours.\newline

We also discussed decompactifications limits, the $R_{B}\to\infty$
one implies $A_{11}\to 0$. In that limit the spectrum of the MIM2
hamiltonian reduces to the ten-dimensional $(p,q)$ type $IIB$
spectrum. This is in agreement with \cite{russo}, where the
$A_{11}\to 0$ limit of the supermembrane was analysed.\newline

Finally we introduced the MIM2 theory compactified on a T-dual
torus. In the same way as the $(p,q)$ $IIB$ string states can be
identified on the MIM2 there is multiplet of $(p,q)$
non-perturbative states associated or the MIM2 with a corresponding
 T-dual $SL(2,Z)$ symmetry. Although the relation of these states to
lower dimensional bound states has not yet been determined, we
expect this symmetry to be the conjectured  type $IIA$ $SL(2,Z)$.\newline
  The KK states of the supermembrane may
be associated to a dynamical tension,  from type IIB perspective it
is seen as an additional dimension as pointed out by
\cite{azcarraga},\cite{townsend}. The stability of the  $SL(2,Z)$
dyonic strings is a natural consequence of the quantum stability of
the MIM2, since it is contained in its spectrum and it fully agrees
with the results of stability of these bound states of strings
founded in \cite{schw2} and \cite{roy}. The full symmetry in fact
encodes also the transformation of the radii, and the gauge field in
a nontrivial way. This type of S-duality is suprising and may be
constitute an indication of a putative lift of the degeneracy of the
radii.

We are dealing with T-duality and S-duality at the non-perturbative
level. The T-duality which we consider represents an extension of
the Buscher rules. In \cite{hulltownsend} they conjectured in the
context of supergravity the existence of some simmetries: the
nonperturbative version of T-duality $O(n,n,Z)$ and S-duality
 $SL(2,Z)$ which in low dimensions could be immersed a unified discrete
symmetry group  $U$. Some cases like the $N=8,4$ supergravity
analysis were explicitly considered.  The low effective action of
the MIM2 should present this discrete symmetry $U$ in lower
dimensions. In fact, in 4D it has $N=1$ supersymmetries and so the
particular discrete groups proposed do not hold on this analysis.
However, it has been pointed out that the $N=1$ in 4D supergravity
the discrete U-group is associated to the symplectic groups
\cite{dauria}. The action of the supermembrane in 4D with $N=1$
supersymmetry has been recently obtained and it contains as a
symmetry group $Sp(6,Z)$. One could think that the $Sp(6,Z)$ may be
related to  the full discrete group of U-duality for $N=1$
supersymmetry however further study to determine it  would be
needed.

 Associated to the MIM2 theory there
 exists a $SL(2,Z)$ multiplet of conserved charges $(q,p)$ whose field
  strengths are the $H_{3}, F_{3}$ on type IIB string theory.
From the type IIB supergravity effective actions this may represent
a nonperturbative origin of these set of fluxes, -a well known open
question-. Fluxes, \cite{GKP}, have become very important in the
search for realistic compactifications. They are able to smooth
singularities, give masses to moduli as well as allow to break
supersymmetry in a controlled manner.
 At present their values are  completely arbitrary,  and so far there has been lacking
 an explanation of its origin at a noneffective theory. Our results indicate that at
 least a subset of them  would be related to the winding of the MIM2 and consequently
 can acquire a more intrinsic meaning since for a particular theory  with a fixed value
 of the central charge $n$, those values are specified. An analogous reasoning
 can  be in principle applied to the subset of  fluxes associated to the $SL(2,Z)$ on
 the $IIA$ supergravity effective theory.

At the level of D-brane interpretation, the supermembrane with
central charges represents the M-theory lifting of some bound
states.  We have seen that the MIM2 contains the bound states of
dyonic strings of $IIA$ and $IIB$ theories. In \cite{d2-d0} it was
found that the MIM2 in 9D corresponds to  a bundle of D2-D0 states
in the type IIA picture.
 Bound states of Dp-branes are associated to  mixed boundary conditions on the strings
 attached to the (Dp,Dq) branes, \cite{lerda}. In the case of (F,Dp) branes the boundary
 state formalism may also be applied \cite{frau}. It would be
interesting to precise the
 relation between both types of bound states since they are
 examples of nonpertubative corrections in string theory with
 well-developed techniques of computation, see for example
 applications in relation with the $SL(2,Z)$ duality \cite{vandoren}.
\section{Acknowledgements}
We would like to thank  for very fruitful conversations to M.~Frau,
O.~Lechtenfeld, A.~Lerda, P.~Merlatti, H. ~Nicolai, T.~Ortin,
J.~Rosseel and A.~Uranga.
 M.P.G.M. is
partially supported by Dipartimento di Fisica di Universita di
Torino under European Comunity's Human Potential Programme and by
the Italian MUR under contracts PRIN-2005023102 and PRIN-2005024045.
I.M. is funded by Decanato de Investigaciones y Desarrollo
(DID-USB). Proyecto G11, USB, Venezuela. The work of A. R. is
supported by a grant from MPG, Albert Einstein Institute, Germany
and  I.M. and A.R. are also supported by PROSUL, under contract CNPq
490134/2006-08.


\begin{thebibliography}{99}

\bibitem{julia} B. Julia in
{\em Supergravity and Superspace}, Eds. S. W. Hawking and M. Rocek,
Cambridge University Press, 1981.

\bibitem{hulltownsend}C.M. Hull, P.K. Townsend
 {\em Unity of superstring dualities.}
 Nucl.Phys.{\bf B438}:109-137,1995. e-Print: hep-th/9410167

\bibitem{schwarz} J. H. Schwarz {\em An SL(2,Z) Multiplet of Type IIB Superstrings}
Phys.Lett. {\bf B360} (1995) 13-18; Erratum-ibid. {\bf B364}
(1995)252 \\
arXiv:hep-th/9508143


\bibitem{schw2}J. H. Schwarz
{\em Lectures on superstring and M theory dualities: Given at ICTP
Spring School and at TASI Summer School. (Caltech) . CALT-68-2065,}
Nucl.Phys.Proc.Suppl.{\bf 55B}:1-32,1997. hep-th/9607201


\bibitem{witten} E. Witten {\em
Bound states of strings and p-branes}.  Nucl.Phys.{\bf
B460}:335-350,1996. e-Print: hep-th/9510135



\bibitem{boonstra} E. Bergshoeff, H. J. Boonstra, T. Ortin
{\em S duality and dyonic p-brane solutions in type II string
theory.} Phys.Rev.{\bf D53}:7206-7212,1996. hep-th/9508091

\bibitem{tomas} P. Meessen, T. Ortin {\em An Sl(2,Z) multiplet of nine-dimensional
type II supergravity theories.} Nucl.Phys.{\bf B541}:195-245,1999.
hep-th/9806120

\bibitem {roy} S. Roy
{\em Sl(2,z) multiplets of type II superstrings in d < 10.}
Phys.Lett.{\bf B421}:176-184,1998. hep-th/9706165

\bibitem{jabbari}H.
Arfaei, M.M. Sheikh Jabbari {\em Mixed boundary conditions and
brane, string bound states.} Nucl.Phys.{\bf B526}:278-294,1998.
e-Print: hep-th/9709054

\bibitem{Lu}J.X. Lu , S. Roy
{\em (F, D5) bound state, SL(2, Z) invariance and the descendant
states in type IIB / A string theory.}  Phys.Rev.{\bf
D60}:126002,1999. hep-th/9905056


\bibitem{azcarraga} J.A. de Azcarraga, J.M. Izquierdo, P.K. Townsend
 {\em A Kaluza-Klein origin for the superstring tension.}
Phys.Rev.{\bf D45}:3321-3325,1992.

\bibitem{townsend}P.K. Townsend
{\em Membrane tension and manifest IIB S duality. } Phys.Lett.{\bf
B409}:131-135,1997. e-Print: hep-th/9705160


\bibitem{abou} M. Abou-Zeid, B. de Wit, D. L\"ust, H. Nicolai, {\em
Space-time supersymmetry, IIA/B Duality and M Theory} Phys.Lett.{\bf
B466}:144-152,1999. hep-th/9908169


\bibitem{uehara} H. Okagawa, S.
Uehara, S. Yamada {\em (p,q)-string in the wrapped supermembrane on
2-torus: A classical analysis of the bosonic sector.} Phys.Lett.{\bf
B639}:101-109,2006.  hep-th/0603203


\bibitem{uehara2} H. Okagawa, S.
Uehara, S. Yamada {\em (p,q)-string in matrix-regularized membrane
and type IIB duality} arXiv:0708.3484


\bibitem{bst} E. Bergshoeff, E. Sezgin, P.K. Townsend,
{\em Supermembranes and eleven-dimensional supergravity.} Phys.
Lett. {\bf B189}: 75-78, 1987.

\bibitem{bellorin} J. Bellorin, A. Restuccia,
{\em D=11 Supermembrane wrapped on calibrated submanifolds}
Nucl.Phys. {\bf B737} 190-208, 2006. {\tt hep-th/0510259}

\bibitem{bgmr2} L. Boulton, M.P. Garcia del Moral, A. Restuccia
{\em The Supermembrane with central charges: (2+1)-D NCSYM,
confinement and phase transition.}{\tt hep-th/0609054} To appear on
Nucl. Phys.{\bf B}.

\bibitem{gmpr}M.P. Garcia del Moral, J.M. Pena, A. Restuccia
{\em  N=1 4D Supermembrane from 11D.} e-Print:
arXiv:0709.4632[hep-th]

\bibitem{torrealba} I. Martin, A. Restuccia, R. S. Torrealba,
{\em On the stability of compactified D = 11 supermembranes.}
 Nucl. Phys. {\bf B521}: 117-128, 1998. {\tt hep-th/9706090}

\bibitem{mor}I. Martin, J. Ovalle, A. Restuccia,
{\em D-branes, symplectomorphisms and noncommutative gauge theories.}
Nucl. Phys. Proc. Suppl. {\bf 102}: 169-175, 2001; {\em Compactified
D = 11 supermembranes and symplectic noncommutative gauge theories.
} Phys. Rev.{\bf D64}: 046001, 2001. {\tt hep-th/0101236}

\bibitem{dwhn} B. de Wit, J. Hoppe, H. Nicolai, {\em On the quantum mechanics of
supermembranes}. Nucl. Phys. {\bf B305}: 545,1988.


\bibitem{gmr} M.P. Garcia del Moral, A. Restuccia, {\em On the
spectrum of a noncommutative formulation of the D=11 supermembrane
with winding} Phys.Rev. {\bf D66} 045023, 2002. {\tt hep-th/0103261}


\bibitem{bgmmr} L. Boulton, M. P. Garcia del Moral, I.
Martin, A. Restuccia  {\em On the spectrum of a matrix model for the
D=11 supermembrane compactified on a torus with non-trivial
winding.} Class. Quant. Grav. {\bf 19}  2951, 2002. {\tt
hep-th/0109153}


\bibitem{bgmr} L. Boulton, M.P.
Garcia del Moral, A. Restuccia, {\em Discreteness of the spectrum of
the compactified D=11 supermembrane with non-trivial winding}.
Nucl.Phys. {\bf B671} 343-358, 2003. {\tt hep-th/0211047}

\bibitem{br} L. Boulton and A. Restuccia,
{\em The Heat kernel of the compactified D=11 supermembrane with
non-trivial winding}. Nucl. Phys. {\bf B724} 380-396, 2005. {\tt
hep-th/0405216}


\bibitem{dwln} B. de Wit, M. Luscher, H. Nicolai, {\em The supermembrane is unstable}.
Nucl. Phys. {\bf B320}: 135, 1989.

\bibitem{dwpp} B. de Wit, K. Peeters, J. Plefka,
{\em Supermembranes with winding.} Phys. Lett.{\bf B409}: 117-123,
1997. {\it hep-th/9705225}


\bibitem{witten3} E. Witten,
{\em Two dimensional Gravity and intersection theory on moduli
space}. Survey in Diff. Geom. {\bf 1} (1991) 243-310.

\bibitem{stelle} M.J. Duff, T.
Inami, C.N. Pope, E. Sezgin, K.S. Stelle, {\em Semiclassical
Quantization Of The Supermembrane.} Nucl. Phys. {\bf B297}: 515,
1988.

\bibitem{russo}J. G. Russo
{\em Construction of SL(2,Z) invariant amplitudes in type IIB
superstring theory. } Nucl.Phys.{\bf B535}:116-138,1998.
hep-th/9802090

\bibitem{dauria}L. Andrianopoli, M. Bertolini, A. Ceresole, R. D'Auria, S. Ferrara
{\em  N=2 supergravity and N=2 superYang-Mills theory on general
scalar manifolds: Symplectic covariance, gaugings and the momentum
map.}  J.Geom.Phys.{\bf 23}:111-189,1997. hep-th/9605032



\bibitem{GKP} S. B. Giddings, S. Kachru, J. Polchinski
{\em Hierarchies from Fluxes in String Compactifications}
 Phys.Rev.{\bf D66} (2002) 106006, arXiv:hep-th/0105097


\bibitem{d2-d0} M.P. Garcia del Moral, A. Restuccia,
{\em The Supermembrane with central charge as a bundle of D2 - D0
branes.} Institute of Physics Conference Series 2005, Vol {\bf 43},
151. {\tt hep-th/0410288}


\bibitem{lerda} P. Di Vecchia, M. Frau, I. Pesando, S.
Sciuto, A. Lerda, R. Russo
 {\em Classical p-branes from boundary state.}
Nucl.Phys.{\bf B507}:259-276,1997. hep-th/9707068

\bibitem{frau} P. Di Vecchia, M. Frau, A. Lerda, A. Liccardo
    {\em (F,Dp) bound states from the boundary state}
     Nucl.Phys.{\bf B565} (2000) 397-426, arXiv:hep-th/9906214

\bibitem{vandoren} D. Robles-Llana, M. Rocek, F. Saueressig, U. Theis,
S. Vandoren {\em Nonperturbative corrections to 4D string theory
effective actions from SL(2,Z) duality and supersymmetry.}
Phys.Rev.Lett.{\bf 98}:211602,2007. hep-th/0612027


\end{thebibliography}
\end{document}